\documentclass[aps,twocolumn,showpacs,floatsfix]{revtex4}

\usepackage{graphicx,epic,eepic}% Include figure files
\usepackage{bm}% bold math
\usepackage[]{graphics,epsfig,amssymb,color}
\newcommand{\be}{\begin{equation}}
\newcommand{\ee}{\end{equation}}
\begin{document}
%\preprint{JPhysB}
\title{Thermal entanglement and efficiency of the quantum Otto cycle for the su(1,1) Tavis-Cummings system}
%Manuscript Title:\\with Forced Linebreak}% Force line breaks with \\
%\title{Dynamics of dipolar molecular chain in crossed static: Soliton formation}
%Manuscript Title:\\with Forced Linebreak}% Force line breaks with \\

\author{L. Chotorlishvili$^{1,3}$, Z. Toklikishvili$^{2}$,  J. Berakdar$^{1}$ }

\affiliation{1 Institut f\"ur Physik, Martin-Luther Universit\"at
Halle-Wittenberg, Heinrich-Damerow-Str.4
06120 Halle, Germany\\
 2 Physics Department of the Tbilisi State University,
                Chavchavadze av.3, 0128, Tbilisi, Georgia\\
 3 Institut f\"ur Theoretische Physik Universit\"at Heidelberg
Philosophenweg 19, D-69120 Heidelberg,Germany}

\begin{abstract}
The influence of the dynamical Stark shift on the thermal entanglement and the
efficiency of the quantum Otto cycle is studied for the su(1,1)
Tavis-Cummings system. It is shown that the degree of the thermal
entanglement becomes larger as the dynamical Stark shift increases.
In contrast, the efficiency of the Otto cycle is degraded with an increase
of the values of dynamical Stark shift. Expressions for the
efficiency coefficient are derived. Using those expressions we
identify the maximal efficiency of the quantum Otto cycle
from  the experimentally measured values of the dynamical Stark shift
\end{abstract}
\maketitle

\section{Introduction }
     The Jaynes and Cummings (JC) \cite{Jaynes} model is a paradigm   of quantum optics describing  a
     two-level system coupled to a single mode of the radiation field. Due to its relative simplicity
     and relevance to applications in diverse  areas of modern optics such as cavity quantum electrodynamics
     \cite{Schleich,Raimond}
     and quantum state engineering with Josephson junction devices \cite{Makhlin},
     JC model is still  of current  interest. The JC model and its generalizations
     \cite{Joshi,Moya-Cessa, Rybin}
      is particularly useful  for studies  in the field of quantum information processing \cite{Mabuchi}.  A  key issue  in this context
      is the concept of the entanglement and the correlations between quantum states \cite{Amico}. In recent years attention was devoted to  so-called quantum engines \cite{Shevchenko,Maruyama,Buluta,Hanggi} and the relation between the thermodynamics
      and the entanglement \cite{Vedral,Quan}. In particular, two aspects are addressed: The
      relation between  the degree of  the entanglement and the efficiency of quantum engines  \cite{Quan} and the role  of thermal
      fidelity as an indicator  for a  thermal phase transition \cite{Cucchietti,Zanardi}.
      {Generally, studies of thermodynamical effects in nanosized systems are expected to reveal the connection between quantum physics and thermodynamics contributing to the understanding of fundamental physical problems, such as the Maxwell's demon and the universality of the  second law of thermodynamics \cite{Maruyama}. Our focus here is on the  so-called quantum heat engine that produces work based on  quantum effects. For instance, for a quantum cavity one considers  the quantum Otto cycle. This cycle contains two quantum adiabatic and two quantum isochoric parts and therefore preserves the volume of the cavity.}
      As a theoretical model we utilize  the su(1,1) Tavis-Cummings (TC) model
      \cite{Joshi}
      including the dynamical Stark shift (DSS) effect \cite{Moya-Cessa}. This model describes two-photon transitions
       between the ground and  the excited state  via  an intermediate state.  The intermediate state
       can be eliminated from the equations of motion \cite{Alsing,Puri} on the
        cost of introducing a  dynamical Stark shift \cite{Joshi}.  Recently, the influence of DSS on the system's concurrence and fidelity was studied  \cite{Chotorlishvili}.
       The main goal here is to study of the relation between the efficiency of the quantum Otto engine and DSS. We consider a quantum
       thermodynamical cycle for  experimentally accessible  situations. Specifically we consider a scenario where a system, consisting of
        atoms and a quantum cavity, is connected to  two thermal baths one with a
       high and the other with a low temperature. As shown below, there is a direct relation between DSS and
       the efficiency of the quantum
       Otto engine; for a given DSS one can predict the efficiency of the
        quantum Otto engine  and visa versa, i.e. from the evaluated efficiency
       one may infer the values of DSS.
       \section{Thermal concurrence}
       The Hamiltonian of two TC atoms placed in an ideal cavity reads \cite{Joshi}
       \begin{eqnarray}
       \hat{H} &=& \big(\omega_{0}+\xi\hat{a}^{+}\hat{a}\big)S_{1}^{z} + \big(\omega_{0}+\xi\hat{a}^{+}\hat{a}\big)S_{2}^{z} + \omega\hat{a}^{+}\hat{a}+ \nonumber \\
               &+& g\bigg[\big(S_{1}^{+}+S_{2}^{+}\big)\hat{a}^{2}+\big(S_{1}^{-}+S_{2}^{-}\big)\big(\hat{a}^{+}\big)^{2}\bigg],
       \label{eq:hamiltonian}
       \end{eqnarray}
       where $g$ is the coupling constant between the atoms and the
       radiation field of the cavity. $a$  and $a^{+}$  are
the photon annihilation and creation operators.  $\xi$ is the strength of
the Stark shift that results in an intensity dependent transition
frequency. The atomic two-level systems are described by the spin operators
$S^{z}=\frac{1}{2}\sigma_{z}$,
$S^{\pm}=\frac{1}{2}\big(\sigma_{x}\pm i\sigma_{y}\big)$,  where
$\sigma_{x,y,z}$ are Pauli operators. In (\ref{eq:hamiltonian}) we neglect the
kinetic energy  $\frac{P^{2}}{2M}$ of the center of mass
 of the atoms (with momentum $P$ and mass $M$). This is valid as long as
 $\frac{P^{2}}{2M}<<d\sqrt{\frac{\hbar\omega n}{2\varepsilon_{0}V}}$ \cite{Schleich}, where
 $d$ is the atomic dipole moment, $V$ is the volume of the cavity, $\varepsilon_{0}$ is the electric constant
 and $n$ is the number of photons in the cavity.
  Due to the presence of the Stark
shift term the eigenfrequency of the system is not constant anymore.
Usually similar problems arise in non-stationary dynamics of
nonlinear systems, where the frequency shift has a crucial consequences
\cite{Schwab}. Here in a stationary problem we also expect to see
consequences of the frequency shift. As the Stark shift term changes the
system's energy spectrum it should also affect the
thermodynamic characteristics of the quantum cycle. In the limit of strong $g\gg \zeta$ or weak  $g\ll\zeta$ coupling regimes
between radiation field of cavity and atomic subsystem, in the interaction representation Hamiltonian (1)
 splits in two commuting parts  \cite{Chotorlishvili}. However, for moderate  $\zeta\sim g$ we expect to see strong correlation effects.
           As usual, the
 radiation filed is  assumed to be prepared in a coherent state with the
distribution function
$W_{n}^{2}=\frac{\alpha^{2n}}{n!}e^{-\alpha^{2}}$, where
  $\alpha$
is the mean photon number $n$.
 We use the following notation for the relevant Hilbert space vectors
  ($g$ and $e$ stand for the ground and excited state)
 \be |eg,n+2\rangle,~~~|ge,n+2\rangle,
~~~|ee,n\rangle,~~~|gg,n+4\rangle .\label{eq:state}
 \ee
In this basis and for
  the resonant case, i.e. for  $\omega_{0}=2\omega$, we construct density operator of the system. {Since working substance of the Otto engine is formed by two TC atoms we are interested in the density operator of the atomic subsystem. Note that the field is the subsystem with a large number of degrees of freedom and is prepared in a coherent state. Therefore, the state of the field is not influenced by the atom-field coupling interaction.  In particular  the state of the field can never be identified exactly due to its probabilistic nature. Consequently, for quantifying the entanglement between the atoms we should average and trace out the field states which leads to a loss of coherence between the atomic and the  field subsystems. Nevertheless, the atomic states are still correlated and we expect to obtain a nonzero concurrence. In the language of density matrix, the coherence of the atomic states  means non-vanishing off-diagonal matrix elements of the reduced density matrix. In addition to  the connection between the engine efficiency and the thermal concurrence, further important issues to be discussed are the consequences of the system's transition form pure to the mixed state on the engine efficiency.} Using basis vectors (2) after tracing field states for the reduced  density operator
  of the atomic subsystem  $\rho_{a}=Tr_{f}\big(\frac{e^{-\beta H_{int}}}{Z}\big)$  we deduce
 \begin{eqnarray}
  \rho_{11} &=& \frac{1}{Z}\frac{2b^{2}}{\big(a^{2}+4b^{2}\big)}+ \frac{1}{Z}\frac{\big(a^{2}+2b^{2}\big)\cosh\big[\beta\sqrt{a^{2}+4b^{2}}\big]}{a^{2}+4b^{2}}-\nonumber \\
         &\frac{1}{Z}& \frac{a\sinh\big[\beta\sqrt{a^{2}+4b^{2}}\big]}{\sqrt{a^{2}+4b^{2}}},
 \nonumber \\
 \rho_{22} &=& \frac{1}{Z}\frac{2b^{2}}{\big(a^{2}+4b^{2}\big)}+\frac{1}{Z}\frac{\big(a^{2}+2b^{2}\big)\cosh\big[\beta\sqrt{a^{2}+4b^{2}}\big]}{a^{2}+4b^{2}}+\nonumber \\
       &\frac{1}{Z}& \frac{a\sinh\big[\beta\sqrt{a^{2}+4b^{2}}\big]}{\sqrt{a^{2}+4b^{2}}},
 \nonumber \\
\rho_{12}&=&\rho_{21}=\rho_{34}=\rho_{43}=\nonumber \\
&\frac{1}{Z}&\frac{2b^{2}\bigg(-1+\cosh\big[\beta\sqrt{a^{2}+4b^{2}}\big]\bigg)}{a^{2}+4b^{2}},\nonumber \\
\rho_{13}&=&\rho_{14}=\rho_{31}=\rho_{41}=\nonumber\\
&\frac{1}{Z}&\frac{ab\bigg(-1+\cosh\big[\beta\sqrt{a^{2}+4b^{2}}\big]\bigg)}{a^{2}+4b^{2}}\nonumber \\
-&\frac{1}{Z}&\frac{b\sinh\bigg(\beta\sqrt{a^2+4b^2}\bigg)}{\sqrt{a^2+4b^2}},\nonumber\\
\rho_{23}&=&\rho_{32}=\rho_{24}=\rho_{42}=\frac{1}{Z}\frac{ab\bigg(1-\cosh\big[\beta\sqrt{a^{2}+4b^{2}}\big]\bigg)}{a^{2}+4b^{2}}-\nonumber\\
&\frac{1}{Z}& \frac{b\sinh\bigg(\beta\sqrt{a^2+4b^2}\bigg)}{\sqrt{a^2+4b^2}},\nonumber\\
\rho_{33}&=&\rho_{44}=\frac{1}{Z}\frac{a^2+b^2+2b^2\cosh\bigg[\beta\sqrt{a^2+4b^2}\bigg]}{a^2+4b^2}. \nonumber \\
\end{eqnarray}
The partition function reads
 \be \label{eq:rho}
 Z=Tr\big(e^{-\beta
 H}\big)=2\bigg(1+\cosh\big(\beta\sqrt{a^2+4b^2}\big)\bigg)
 .\ee
Here $a=\xi\alpha^2$, $b=g\alpha^2$.
The  eigenvalues of the
Hamiltonian (\ref{eq:hamiltonian}) are $$E_{1}=-\sqrt{a^2+4b^2},\, E_{2}=0,\, E_{3}=0,\,
E_{4}=\sqrt{a^2+4b^2}.$$

 Using (3),(4)
we follow Ref. \cite{Wootters} and  evaluate the
concurrence $C$ according to
$$C=\max\big(0,\sqrt{R_{1}}-\sqrt{R_{2}}-\sqrt{R_{3}}-\sqrt{R_{4}}\big)$$
where the square roots correspond to the eigenvalues of the matrix
$$R=\big(\sigma_{y}\otimes\sigma_{y}\big)\rho^{*}\big(\sigma_{y}\otimes\sigma_{y}\big)\rho.$$
          As we seek  to inspect the influence of the Stark shift
we consider an asymptotic case that corresponds to large values of the
Stark shift $g<<\xi$. In this case  eqs. (3) and (4) simplify and we
obtain
\begin{eqnarray}
\rho&=&\frac{1}{Z}\left(
                    \begin{array}{cccc}
                       e^{-\beta a} & 0           & 0 & 0\\
                       0            & e^{\beta a} & 0 & 0\\
                       0            & 0           & 1 & 0\\
                       0            & 0           & 0 & 1
                    \end{array}
                  \right), \\
R&=&\frac{1}{Z^2}\left(
                   \begin{array}{cccc}
                       e^{-\beta a} & 0           & 0           & 0\\
                       0            & e^{\beta a} & 0           & 0\\
                       0            & 0           & e^{\beta a} & 0\\
                       0            & 0           & 0           & e^{-\beta a}
                   \end{array}
                  \right),
\end{eqnarray}
\begin{equation}
Z=2\big(1+\cosh\beta a\big). \nonumber
\end{equation}
Furthermore, we obtain for the eigenvalues $$R_{1,2}=\frac{e^{\beta a}}{Z^2},\,
R_{3,4}=\frac{e^{-\beta a}}{Z^2},\, R_{1,2}>R_{3,4}$$ and
therefore  $$C=\max\big(0,-2\sqrt{R_{3}}\big)=0.$$ Obviously a large
Stark shift leads to a disentanglement. The same result we obtain
in the case of a small Stark shift. Indeed,  considering $g>>\xi$
we infer from eq. (3) and (4) that
 \be
\rho=\left(\begin{array}{c}\frac{1}{4}~~a_{1}~~a_{2}~~a_{2}\\a_{1}~~\frac{1}{4}~~a_{2}~~a_{2}\\a_{2}~~a_{2}~~\frac{1}{4}~~a_{1}\\
a_{2}~~a_{2}~~a_{1}~~\frac{1}{4} \end{array}\right).\ee
In what follows we introduce the
following notations
$$a_{1}=\frac{1}{2Z}\big(-1+\cosh 2\beta b\big),\,
a_{2}=-\frac{1}{2Z}\sinh 2b\beta,$$ and $$Z=2\big(1+\cosh2b\beta\big).$$
The matrix $R$ and its eigenvalues read
\be
R=\left(\begin{array}{c}\frac{1}{16}-a_{1}^{2}~~~0~~~0~~~\frac{a_{2}}{2}-2a_{1}a_{2}\\0~~~\frac{1}{16}-a_{1}^{2}~~~\frac{a_{2}}{2}-2a_{1}a_{2}~~~0\\
0~~~\frac{a_{2}}{2}-2a_{1}a_{2}~~~\frac{1}{16}-a_{1}^{2}~~~0\\\frac{a_{2}}{2}-2a_{1}a_{2}~~~0~~~0~~~\frac{1}{16}-a_{1}^{2}\end{array}\right)\ee
$R_{1,2}=\frac{e^{6\beta b}}{\big(1+e^{2\beta
b}\big)^2}$,$R_{3,4}=\frac{e^{2\beta b}}{\big(1+e^{2\beta
b}\big)^2}$,
 and hence
 \be C=\max\big(0,-2\sqrt{R_{3}}\big)=0 \ee
 This Result
is quite simple and evident. The value of DSS is the key factor since
for both limiting cases the system is disentangled. To consider
arbitrary values of DSS we introduce the following notations
\begin{eqnarray}
&x&=\frac{2b}{a}=\frac{2g}{\xi}, ~~ y=\beta a=\frac{\xi\alpha^2}{T},\nonumber \\
&b_{1}& =\frac{1}{2}\frac{x \cosh\big[y\sqrt{1+x^2}\big]}{1+x^2},\nonumber \\
&
b_{2}& =\frac{1}{2}\frac{x \sinh\big[y\sqrt{1+x^2}\big]}{1+x^2}.
\end{eqnarray}
and rewrite the density matrix in the form
\begin{widetext}
\be\label{eq:density}
\rho=\frac{1}{Z}\left(\begin{array}{c}1+xb_{1}+\frac{2}{x}(b_{1}-b_{2})~~~xb_{1}~~~(b_{1}-b_{2})~~~(b_{1}-b_{2})\\
xb_{1}~~~1+xb_{1}+\frac{2}{x}(b_{1}+b_{2})~~~-(b_{1}+b_{2})~~~-(b_{1}+b_{2})\\
(b_{1}-b_{2})~~~-(b_{1}+b_{2})~~~(1+xb_{1})~~~xb_{1}\\
(b_{1}-b_{2})~~~-(b_{1}+b_{2})~~~xb_{1}~~~1+xb_{1}\end{array}\right),
\ee
$$Z=2\bigg(1+\cosh y\sqrt{1+x^2}\bigg).$$
  Consequently,  the matrix
  $R=\big(\sigma_{y}\otimes\sigma_{y}\big)\rho^{*}\big(\sigma_{y}\otimes\sigma_{y}\big)\rho$ reads
  \be\label{eq:R}
  R=\frac{1}{Z^2}\left(\begin{array}{c}\frac{(1+2b_{1}x)(x+2(b_{1}-b_{2}))}{x}~~~4b_{1}(-b_{1}+b_{2})~~~\frac{2b_{1}(2b_{1}+x)-4b_{2}^{2}}{x}~~~\frac{2(x+2(b_{1}-b_{2}))(b_{1}-b_{2})}{x}\\
  -4b_{1}(b_{1}+b_{2})~~~\frac{(1+2b_{1}x)(x+2(b_{1}+b_{2}))}{x}~~~\frac{-2(b_{1}+b_{2})(x+2(b_{1}+b_{2}))}{x}~~~\frac{-2b_{1}(2b_{1}+x)+4b_{2}^2}{x}\\
  2b_{1}(1+2b_{1}x)~~~-2(4b^2_{1}x+(b_{1}+b_{2}))~~~\frac{(1+2b_{1}x)(x+2(b_{1}+b_{2}))}{x}~~~4b_{1}(b_{1}-b_{2})\\
  2(b_{1}-b_{2}+2b^{2}_{1}x)~~~-2b_{1}(1+2b_{1}x)~~~4b_{1}
  (b_{1}+b_{2})~~~\frac{(1+2b_{1}x)(x+2(b_{1}-b_{2}))}{x}\end{array}\right).
  \ee
\end{widetext}

The eigenvalues of the Matrix (\ref{eq:density}) and consequently the
thermal concurrence is a function of two parameters: The first
parameter is the relation between the cavity-atom coupling constant
and the Stark shift $x=\frac{2g}{\xi}$. The second one describes the
temperature dependence of the  concurrence $y=\frac{\xi\alpha^2}{T}$.
\begin{figure}[t]
 \centering
  \includegraphics[width=8cm]{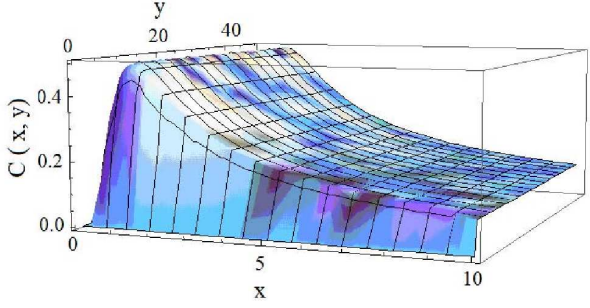}
  \caption{Thermal concurrence as a function of the parameter $x=\frac{2g}{\xi}$,$y=\frac{\xi\alpha^2}{T}$.
  For large and
  small  of Stark shifts the system is not entangled. A maximal concurrence value is
  achieved for an intermediate Stark shift.} \label{Fig:1}
\end{figure}
  Fig.1 is deduced from the calculations using (\ref{eq:R}).
  From this figure we conclude  that the
dependence on the first parameter is more relevant for  the thermal
concurrence.
Therefore, the relation between the cavity-atom coupling
constant and the Stark shift is the key parameter. In the limiting cases
$\xi>>g$, and $\xi<<g$, as was shown before analytically, the  thermal
concurrence is zero.
The concurrence increases linearly with the parameter $g/\xi$  up to the maximal value $C_{max}(g/\xi)\approx 0.5$.
This fact can be explained analytically. Considering in eqs. (3) and (4) $g^{2}/\xi^{2}$ as a small parameter
 and $a\beta$ as a large parameter and retaining only the first order terms we find that $R$ matrix reduces to

 \begin{eqnarray}R&=&\left(
                   \begin{array}{cccc}
                       0            & 0           & 0           & 0\\
                       -2g^{2}/\xi^{2}            & 2g^{2}/\xi^{2} & 2g/\xi          & 0\\
                       0            & 0           & g^{2}/\xi^{2} & 0\\
                       0            & 0           & 2g^{2}/\xi^{2}           & 0 \nonumber
                   \end{array}
                  \right).
\end{eqnarray}

 Consequently the expression for the  concurrence $~~C(g/\xi)=max(0, g/\xi(\sqrt{2}-1))$ is in line with
  the numerical result.  In the opposite case, i.e. for $\xi<g$, and
  $\frac{b\beta}{(g/\xi)^{2}}\geqslant1$ we infer for the concurrence a square root decay
   \begin{eqnarray}
   C(g/\xi)&=&max (0, \sqrt{\frac{2b\beta}{(g/\xi)^{2}}-1+
   \sqrt{\frac{2b\beta}{(g/\xi)^{2}}-1}}\nonumber\\
   && -
   \sqrt{\frac{2b\beta}{(g/\xi)^{2}}-1-
   \sqrt{\frac{2b\beta}{(g/\xi)^{2}}-1}}),
   \end{eqnarray}
    which is also in a good agreement with the numerical results.
    Namely in the regime  of weak coupling between atom and cavity concurrence is linear function of $g/\xi$.
    In the regime of strong coupling we have square root law for decay of concurrence.
\section{Efficiency of quantum Otto engine and the dynamical Stark shift}
Having defined the thermal concurrences for our system we focus now
on the relation between the values of the Stark shift, the
concurrence and the efficiency of the quantum Otto engine.  A quantum Otto
engine with two two-level atoms as a working substance was
proposed and discussed in details in \cite{Quan}.
Here we recall very
briefly  the main  ideas: Two TC atoms are
placed in an ideal lossless cavity which is  connected to two thermal
baths with different temperatures $\beta_{L}=\frac{1}{T_{L}}$,
$\beta_{H}=\frac{1}{T_{H}}$. The system performs a quantum
Otto cycle consisting of two adiabatic and two isochoric parts.
The heat exchange with the thermal reservoir changes the population of the
levels $P_{ij}~~~(i=1,2,3,4.j=1,2)$ (isochoric part), while during the
adiabatic parts the performed work changes the structure of the energy terms
$E_{ij}~~~(i=1,2,3,4.j=1,2)$ (due to the change of the atom-cavity
coupling constant \cite{Wang}).
\begin{figure}[t]
 \centering
  \includegraphics[width=8cm]{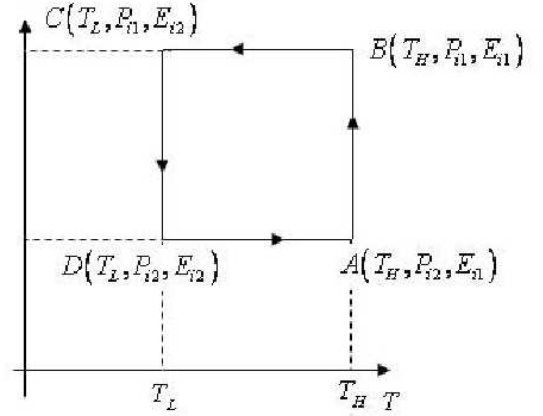}
  \caption{1)$A\rightarrow B$ system is connected with hot reservoir $T_{H}$, structure of energy levels is conserved
  $E_{i1}$ while populations are changed  from $P_{i2}~~(i=1,2,3,4)$ to $P_{i1}~~(i=1,2,3,4)$-Work performed is zero. Isochoric part\\
  2) $B\rightarrow C$ Adiabatic part,  level populations is conserved  $P_{i1}~~(i=1,2,3,4)$.While energy levels are changed from $E_{i1}~~(i=1,2,3,4)$-to
$E_{i2}(i=1,2,3,4)$\\
3) $C\rightarrow D$ is the reverse process of  $A\rightarrow B$\\
4) $D\rightarrow A$ adiabatic part. Populations are conserved
$P_{i2}(i=1,2,3,4)$ energy levels changed $E_{i2}(i=1,2,3,4)$-from
to $E_{i1}(i=1,2,3,4)$} \label{Fig:2}
\end{figure}
{For $P_{ij}$ the first index $i$ enumerates the level population probabilities, while the second index denotes the probabilities before or after the heat transfer from the thermostat to the system. For $E_{ij}$ the first index describes the energy levels while the second defines the energy spectrum before or after performing work (i.e., the performed work changes the structure of the energy spectrum as was stated above.)}  The  quantum Otto cycle is shown schematically in
Fig.2.
 Following the standard notations, e.g. as in \cite{Quan}, we write \be
\label{eq:E}
E_{i1}=\left\{\begin{array}{ll}E_{11}=-\lambda_{1};\\E_{21}=0;\\E_{31}=0;\\E_{41}=\lambda_{1};\end{array}\right.~~~~~
E_{i2}=\left\{\begin{array}{ll}E_{11}=-\lambda_{2};\\E_{21}=0;\\E_{31}=0;\\E_{41}=\lambda_{2};\end{array}\right.
\ee

\begin{eqnarray}
&P_{i1}&=\left\{\begin{array}{ll}P_{11}=\exp(+\beta_{H}\lambda_{1})/Z_{H};\\P_{21}=\frac{1}{Z_{H}};\\P_{31}=\frac{1}{Z_{H}};\\P_{41}=\exp(-\beta_{H}\lambda_{1})/Z_{H};\end{array}\right.\nonumber \\
&P_{i2}&=\left\{\begin{array}{ll}P_{12}=\exp(+\beta_{L}\lambda_{2})/Z_{H};\\P_{22}=
\frac{1}{Z_{L}};\\P_{32}=\frac{1}{Z_{L}};\\P_{41}=\exp(-\beta_{L}\lambda_{2})/Z_{L}.\end{array}\right.
\label{eq:P}
\end{eqnarray}
 Here
 $$\lambda_{1}=\alpha^2\sqrt{\xi^2+4g_{1}^{2}},
\lambda_{2}=\alpha^2\sqrt{\xi^2+4g_{2}^{2}},  a=\alpha^2\xi,$$
 $$Z_{H}=2(1+\cosh\beta_{H}\lambda_{1}), $$ and
$$Z_{L}=2(1+\cosh\beta_{H}\lambda_{2}).$$ Using (14),(15) and the
standard definitions of the thermodynamic quantities such as the energy of the system $E$, and
the transferred heat $Q$ as well as the work performed $A$ we find
\begin{eqnarray}\label{eq:Q}
&&E=\sum\limits_{i=1}^{4}E_{ij}P_{ij}\\
&&dE=\sum\limits_{i=1}^{4}E_{ij}dP_{ij}+\sum\limits_{i=1}^{4}P_{ij}dE_{ij},
\nonumber\\
&&
\left\{\begin{array}{ll}dQ=\sum\limits_{i=1}^{4}E_{ij}dP_{ij},\\dA=\sum\limits_{i=1}^{4}P_{ij}dE_{ij},\end{array}\right.
\\
&Q_{H}&=Q_{AB}= \int\limits_{A}^{B}\sum\limits_{i=1}^{4}E_{i1}dP_{ij}\nonumber\\
&=&\sum\limits_{i=1}^{4}E_{i1}(P_{i1}-P_{i2})=
\nonumber
\\&&=2\lambda_{1}\bigg(\frac{\sinh\beta_{L}\lambda_{2}}{Z_{L}}-\frac{\sinh\beta_{H}\lambda_{1}}{Z_{H}}\bigg)\nonumber
\\
&Q_{L}&=-Q_{CD}=-\int\limits_{C}^{D}\sum\limits_{i=1}^{4}E_{i2}dP_{ij}\nonumber\\
&=&-\sum\limits_{i=1}^{4}E_{i2}(P_{i2}-P_{i1})=
\\&&=2\lambda_{2}\bigg(\frac{\sinh\beta_{H}\lambda_{1}}{Z_{H}}-\frac{\sinh\beta_{L}\lambda_{2}}{Z_{L}}\bigg)=\frac{\lambda_{2}Q_{H}}{\lambda_{1}}\nonumber
\end{eqnarray}
Finally for the efficiency coefficient we obtain
\begin{eqnarray}
&&\eta=\frac{Q_{H}-Q_{L}}{Q_{L}}=1-\frac{\lambda_{2}}{\lambda_{1}}=1-\frac{\sqrt{1+(g_{2}/\xi)^2}}{\sqrt{1+(g_{1}/\xi)^2}}
\\ && g_{1}=g+\delta g, g_{2}=g. \nonumber
\end{eqnarray}

For the small values of the Stark shift we arrive at the  conclusions
$$\eta(\xi<<g)=\frac{Q_{H}-Q_{L}}{Q_{H}}=\frac{\delta g}{g}+\bigg(\frac{\delta
g}{g}\bigg)^2+\ldots,$$
while for large shift we deduce
$$\eta(\xi>>g)=\frac{Q_{H}-Q_{L}}{Q_{H}}\approx\frac{(\delta
g)g}{\xi^2},$$
$$\eta(\xi>>g)<\eta(\xi<<g).$$

Efficiency of Otto engine $\eta(\xi)$, (20) has a maximum for zero values of the Stark shift $$\eta_{max}=\eta(\xi=0)=1-g/(g+\delta g),$$
 when thermal concurrence is zero. Therefore we can argue that quantum correlations in the atomic subsystem hinders realization of
 quantum Otto cycle with maximal efficiency.

\section{Conclusions}  The purpose of the present paper has been  to study  the
influence of DSS on the thermal entanglement and  the efficiency of the
quantum Otto cycle. For this purpose we considered the su(1,1)
Tavis-Cummings system and showed that the degree of the thermal entanglement
follows the values of the DSS. In particular, for vanishing  DSS, i.e.  $\xi=0$ the
system is disentangled, and the degree of the thermal entanglement
increases with  DSS.  The system  becomes  disentangled again only in
the limit $\xi\gg g$, i.e.  for a  weak atom-cavity
coupling. The efficiency of the quantum Otto
cycle is maximal for a small values of DSS
$\eta(\xi<<g)>\eta(\xi>>g)$, i.e. when system is disentangled.
 On the other hand,  $\eta$ decreases
as  $\eta(\xi)\approx 1/\xi^{2}$. Using the
asymptotic expressions for the efficiency coefficient  (20), one may
 identify the maximal efficiently of the quantum Otto cycle
from the experimentally measured DSS. {For $\xi<<g$, the efficiency is $\eta\sim\delta g/g $, where $\delta g $ is the amendment of the atom cavity coupling constant due to the performed work. If $\xi>>g$ then we obtain for the efficiency  $\eta\sim(\delta g) g/\xi^{2}$. For a given value of the DSS, one can quantify the thermal concurrence as well. Again the key point is the relation $\xi/g$. For $\xi/g<1$, the concurrence is given in an explicit analytical expression $C(g/\xi)$ see Eq. (14). For the opposite case $\xi/g>1$ we refer to the expression above.
In the strong coupling limit $\xi<<g$ or large DSS  $\xi>>g$  system is disentangled. From the physical point of view this result is expectable. In both cases the Hamiltonian of the system (1) can be presented as the sum of two commuting parts $\hat{H}_{int}=\hat{H}_{1}(s_{1}^{z},s_{1}^{\pm})+\hat{H}_{2}(s_{2}^{z},s_{2}^{\pm})$, $[\hat{H}_{1},\hat{H}_{2}]=0$. Therefore, the reduced density matrix of the atomic subsystem is a separable $\rho_{a}=\rho_{1}\oplus\rho_{2}$. In the language of thermal entanglement both limits are identical (entanglement is zero), while the engine works with the maximal efficiency only for $\xi<<g$. Therefore, we can conclude that a zero entanglement is essential but not sufficient a criteria to reach  the maximal efficiency. Comparing Eq. (3) with the Eq. (5) we see that with the increase of DSS the system performs a transition from a  pure state to  a mixed state where the non diagonal  elements of the density matrix are zero Eq.(5).  The system produces maximal efficiency for $\xi<<g$, being in the pure state Eq. (8).  Therefore, we can conclude that for  a maximal efficiency  the working substance should be in a pure coherent state and the transition to the mixed state degrades the efficiency of the quantum engine.}

\textbf{Acknowledgments}  The financial support by the Deutsche
Forschungsgemeinschaft (DFG) through SFB 762, the  Grant No.
KO-2235/3 and STCU Grant No.~5053 is gratefully acknowledged.

\end{document}